\documentclass[%
 reprint,
 amsmath,amssymb,
 aps,
pra,
]{revtex4-2}

\usepackage{graphicx}
\usepackage{dcolumn}
\usepackage{bm}
\usepackage{tikz}
\usepackage{tikz-feynman}
\usepackage{hyperref}
\usepackage{subfigure}

\begin{document}

\preprint{APS/123-QED}

\title{Numerical validation of an ultracold Hubbard quantum simulator} 

\author{Ben Currie}
\author{John Sturt}
\author{Evgeny Kozik}%

\affiliation{%
 Department of Physics, King’s College London, Strand, London WC2R 2LS, United Kingdom
}%

\date{\today}

\begin{abstract}

We apply the formally exact Diagrammatic Monte Carlo (DiagMC) method to probe the unprecedentedly low-temperature regime recently achieved in an ultracold-atom quantum simulation of the 2D Hubbard model \href{https://www.nature.com/articles/s41586-025-09112-w}{[Xu \textit{et al.}, Nature \textbf{642}, 909 (2025)]}. Computing the experimentally measured observables directly in the thermodynamic limit with \textit{a priori} control of systematic errors, we find striking agreement with the experimental data across all accessible temperatures — including the lowest, where existing numerical benchmarks show significant deviations. This validates the quantum simulator’s control over systematic errors in this challenging regime and delivers unbiased benchmarks for future method development.
Our results demonstrate that classical algorithms remain competitive with state-of-the-art analogue quantum simulators, and emphasise the importance of controlled numerical methods for continuing the development of these experiments.

\end{abstract}

\maketitle

Analogue quantum simulations of condensed matter systems, such as cold atoms in optical lattices, have emerged as a powerful platform for simulating and probing many-body quantum systems \cite{Bloch2008Review, Ketterle2008Review, Bloch2017Review}.
They offer pristine and highly tunable realisations of widely studied models of condensed matter physics, free from many of the complexities of their solid state counterparts, providing a unique route to understand exotic strongly correlated states of matter.
They are thus an important tool for understanding and harnessing quantum many-body effects that are crucial for quantum technologies, such as 
quantum computing \cite{Saffman2010,Ebadi2021}. Moreover, they are arguably best-placed to demonstrate quantum advantage in the near-term, by addressing classically intractable problems in quantum many-body physics \cite{Daley2022}.

Fermionic atoms in optical lattices provide an excellent environment for simulating toy models of  electrons moving and interacting in solids \cite{Esslinger2010,Jordens2008,GreinerSpin2016,Cheuk2016,Cocchi2016,Tarruell2018, Taie2022,Pasqualetti2024}. A key example is the fermionic two-dimensional Hubbard model on a square lattice, whose Hamiltonian is 
\begin{equation}    H = -t \sum_{\langle ij\rangle,\sigma} c^\dagger_{i,\sigma} c_{j,\sigma} +c^\dagger_{j,\sigma} c_{i,\sigma} -\mu\sum_{i,\sigma} n_{i,\sigma} + U \sum_{i}n_{i\uparrow}n_{i\downarrow} ,
\end{equation} 
with hopping $t$ between nearest-neighbour sites $\langle ij\rangle$, chemical potential $\mu$, interaction strength $U$ and local-occupation $n_{i,\sigma} = c^\dagger_{i,\sigma} c_{i,\sigma}$ for the spin-component $\sigma$. Despite its simplicity, it is widely believed that this model hosts a variety of exotic quantum states -- including those present in the high temperature cuprate superconductors \cite{Lee2006MottInsulator,Proust2019CuprateReview,Scalapino2012review,Arovas2022TheHubbardModel}  -- making it of fundamental importance in modern condensed matter physics. While classical numerical methods have and are continuing to establish a great deal of the Hubbard model's properties \cite{SimonsCollab2015, EK-crossover, MultiMessenger2021}, key questions about the possible low-temperature phases -- and its ability to capture high-temperature superconductivity in particular -- remain unanswered.
Quantum simulations of the Hubbard model thus offer a promising path to discovering new strongly-correlated phases, and are becoming an instrumental part of achieving the long-standing goal of determining its low-temperature phase diagram. In this vein, they have recently been used to realize and probe states with antiferromagnetic order \cite{Mazurenko2017,Boll2016}, charge order \cite{Hirthe2023,Bourgund2025}, bad metallic transport \cite{Brown2019,Xu2019}, string order in one \cite{Hilker2017}  and two \cite{Chiu2019} dimensions, the pseudogap state \cite{chalopin2024,Kendrick2025} and spin transport \cite{Nichols2019}. 

Despite these successes, quantum simulators have until now been limited to temperatures that, in solid-state terms, lie far above room temperature. In a landmark achievement, Xu \textit{et al.}~\cite{Xu2025} have, for the first time, driven the Hubbard model deep into its ultra-low-temperature regime — a breakthrough that provides direct access to its most elusive and long-sought quantum phases. Remarkably, at half-filling, they infer temperatures as low as $T/t=0.05^{0.06}_{-0.05}$ in systems of approximately 340 atoms. 
Combined with a Hubbard interaction of $U/t=8$ and finite doping -- a detuning of the lattice occupancy from one particle per site -- these conditions are directly relevant for understanding high-temperature superconductivity in the cuprates. This is also the regime most challenging for state-of-the-art numerical methods. However, reliable theoretical predictions are essential for calibrating and validating such experiments. 
Xu \textit{et al.} report a significant discrepancy -- well beyond the stated error bars -- between their measurements and calculations of the spin-spin correlation function using the state-of-the-art but approximate constrained-path auxiliary-field quantum Monte Carlo (CP-AFQMC); the only numerical method that could be applied in this regime. This raises the question of whether systematic errors are fully accounted for in the experiment, the calculations, or both, and underscores the critical importance of validating the experimental data for further progress.

In this paper, we apply the formally exact Diagrammatic Monte Carlo (DiagMC) method~\cite{Prokofev1998, VanHoucke2010, Kozik2010} supplemented by the combinatorial summation (CoS) algorithm~\cite{Kozik2024} to compute the observables measured and calculated in Ref.~\cite{Xu2025}. Our results, obtained directly in the thermodynamic limit with \textit{a priori} control of systematic errors, demonstrate excellent agreement with the experimental data across all temperature regimes accessed in Ref.~\cite{Xu2025}, including the lowest temperatures where the CP-AFQMC predictions begin to deviate from the experimental data. 
This confirms the control of systematic errors in the experiment and establishes its ability to faithfully simulate the Hubbard Hamiltonian in this low-temperature regime. At the same time, we demonstrate that CP-AFQMC captures local and short-range correlations well down to the lowest temperatures, while significant discrepancies appear for spin correlations beyond one lattice spacing. At higher temperatures ($T/t=0.25$), our data agrees with unbiased determinantal quantum Monte Carlo (DQMC) results also reported in Ref.~\cite{Xu2025} — as well as with the experiment — but features significantly smaller error bars. Overall, our results indicate that classical algorithms have not yet been surpassed by quantum simulators and suggest that DiagMC provides a powerful tool for validating cutting-edge cold-atom experiments.

A key probe of the antiferromagnetic (AFM) correlations is provided by the real-space equal-time spin-spin correlation function, defined as
\begin{equation}
    C_S(r) = \frac{1}{S^2}(\langle S_z(r) S_z(0)\rangle - \langle S_z\rangle^2)
\end{equation}
where $S_z(r)=\frac{1}{2}(n_\uparrow(r)-n_\downarrow(r))$ is the $z$-component of the spin operator at site $r$, and $S = 1/2$ is the total electron spin.
In their work, Xu \textit{et al.} probe $C_S(r)$ by means of site-resolved imaging of their system and the independent removal of each spin species from the lattice, allowing them to map the spin correlations onto charge correlations \cite{GreinerSpin2016}.
At low temperatures, AFM correlations manifest as strong site-to-site sign alternations in 
$C_S(r)$, and its substantial temperature dependence makes it an ideal quantity for thermometry.
At half-filling, $C_S(r)$ was computed by numerically exact DQMC and AFQMC, and subsequently translated to staggered magnetization to estimate the experimental temperature, yielding the value $T/t = 0.05^{+0.06}_{-0.05}$.
In the more challenging regime of finite-doping, where DQMC is prohibitively expensive, the authors instead use a  comparison of CP-AFQMC and experimental data for the nearest-neighbour spin-correlations $C_S(|r|=1)$.
The CP-AFQMC suggests that $C_S(|r|=1)$ saturates for temperatures $T/t\lesssim 0.1$, with the saturation value in good agreement with the experimental data, giving a consistent temperature estimate to that found at half filling $T/t\lesssim 0.1$. 
On the other hand, a significant discrepancy between the experiment and CP-AFQMC is reported at longer distances in the doped case, which raises questions over the accuracy of this temperature estimate at finite doping.

We compute the spin-spin correlator in the doped case directly in the thermodynamic limit (TDL) and at finite temperature $T$ in the DiagMC framework~\cite{Prokofev1998, VanHoucke2010, Kozik2010}. Applied to the 2D Hubbard model, DiagMC has provided key insights into the nature of the pseudogap phase \cite{EK-alpha,FedorOriginFate2024}, the metal-to-insulator crossover \cite{EK-crossover,EK-spin-charge} and the entropy in the non-Fermi-liquid regime \cite{Lenihan2021}. In this approach, an arbitrary physical observable $A$ is reconstructed---with controlled precision---from its Taylor series $A=\sum_{n=0}^N a_n U^n$ in the interaction strength $U$ up to some maximal diagram order $N$, whereby $a_n$ is computed numerically exactly as a sum of all Feynman diagrams with $n$ interaction lines. In our calculations, all diagram integrands of a given order $n$ are summed deterministically using the recently-introduced CoS algorithm~\cite{Kozik2024}, whose efficiency at the diagram orders accessed here is comparable to that of the widely used CDet method~\cite{Rossi2017}. The integration over the positions of the interaction vertices in space-imaginary-time is then performed via stochastic sampling, yielding the exact value of $a_n$ with a known statistical error bar.

We extrapolate the series to infinite diagram order using the methods of Pad\'e, Dlog Pad\'e~\cite{Baker1961} and Integral Approximants (sometimes called Differential Approximants)~\cite{hunter1979IA}, in which a sequence of approximants are constructed based on the Taylor series coefficients alone, and are used to reconstruct the function---and evaluate the associated reconstruction error---at the Hubbard interaction of choice. The discrepancy between different approximants yields the total systematic error of the extrapolation, as explained in Ref.~\cite{Simkovic2019}. More details of the extrapolation techniques can be found in the supplementary material.

At low temperatures and at half filling, the long (but finite, due to the Hohenberg-Mermin-Wagner theorem) AFM correlation length manifests in the diagrammatic series as a complex conjugate pair of singularities close to the positive real axis~\cite{Lenihan2021}. They pinch the real axis as the temperature is lowered and the correlation length grows, making the extrapolations extremely challenging. However, at finite doping $\gtrsim 10\%$, the singularities move further from the real axis, allowing precise reconstruction of the result from the series. This is our regime of interest, where the discrepancy between numerical data and experiment is most pronounced.

\begin{figure}[h]
\centering
   
\includegraphics[width=0.5\textwidth,trim={3 0 0 10},clip]{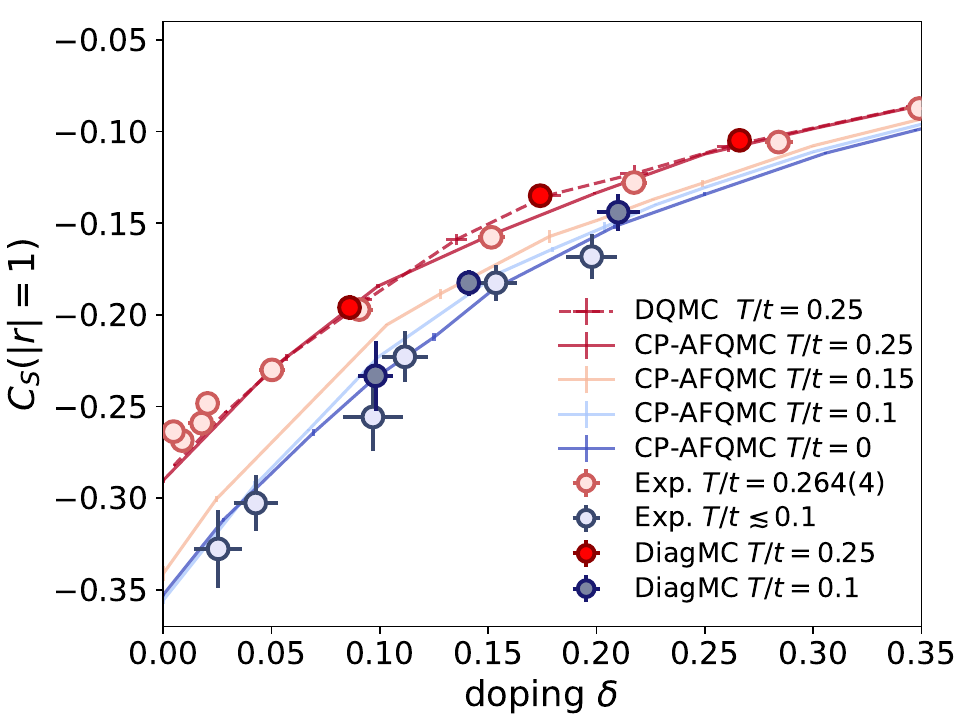}      

\caption{Nearest neighbour spin-spin correlations at $U/t = 8$ and temperatures $T/t=0.1$ and $T/t=0.25$ computed with DiagMC, compared with the experiment of Ref.\cite{Xu2025}, determinantal quantum Monte Carlo (DQMC), and constrained-path auxiliary field quantum Monte Carlo (CP-AFQMC). 
}    
\label{fig:d_10}
\end{figure}

Our result for the nearest neighbour spin-spin correlator at distance $r = (1,0)$ is shown as a function of doping in Fig.~\ref{fig:d_10} for temperatures $T/t=0.25$ and $T/t=0.1$, and compared to the experimental values and numerical data of Ref. \cite{Xu2025}. For both temperatures, the correlator displays a substantial doping dependence, with a maximum at half-filling where antiferromagnetic correlations are strongest. At the higher temperature, our thermodynamic limit data agrees extremely well with DQMC on an $8 \times 8$ system, suggesting an absence of finite size error at this distance and temperature. The lower experimental temperature was reported as $T/t\lesssim 0.1$, and our numerical data at $T/t=0.1$ agrees well with the experimental data. Moreover, the CP-AFQMC data, with which our data shows excellent agreement at both temperatures at this distance, displays a saturation to the zero temperature regime already at $T/t\approx 0.1$.

\begin{figure}[h]
\centering    
   
\includegraphics[width=0.5\textwidth,trim={10 0 0 10},clip]{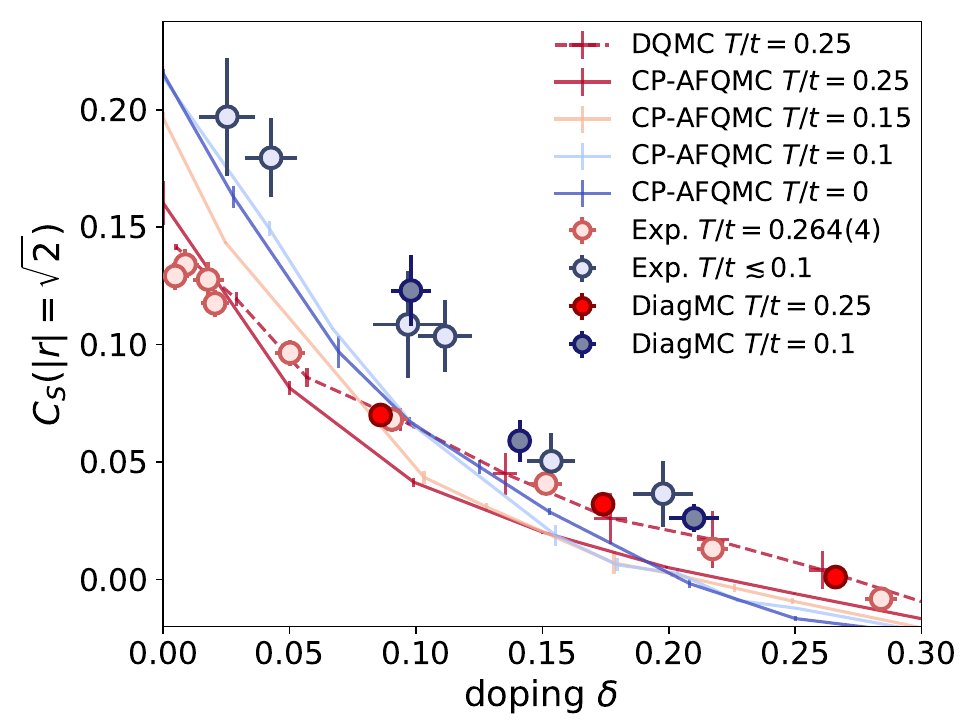}      

\caption{Next-nearest-neighbour diagonal spin correlations correlations at $U/t = 8$ and temperatures $T/t=0.1$ and $T/t=0.25$ computed with DiagMC, compared with the experiment of Ref.\cite{Xu2025}, determinantal quantum Monte Carlo (DQMC), and constrained-path auxiliary field quantum Monte Carlo (CP-AFQMC).}    
\label{fig:d_11}
\end{figure}

Our main result is the diagonal next-nearest-neighbour spin-spin correlator at distance $d=(1,1)$, shown in Fig.~\ref{fig:d_11}.
At the higher temperature, our data agrees well both with the experiment and the determinantal quantum Monte Carlo (DQMC) obtained for system size $8\times8$. On the other hand, we find substantial disagreement with the CP-AFQMC data at this distance and temperature, which systematically underestimates the magnitude of the correlations. 
This discrepancy persists as the temperature is lowered to $T/t=0.1$, where once again our data shows excellent agreement with the experiment, and where unbiased DQMC data is not available.

\begin{figure}[h]

\centering    

\includegraphics[width=0.5\textwidth]{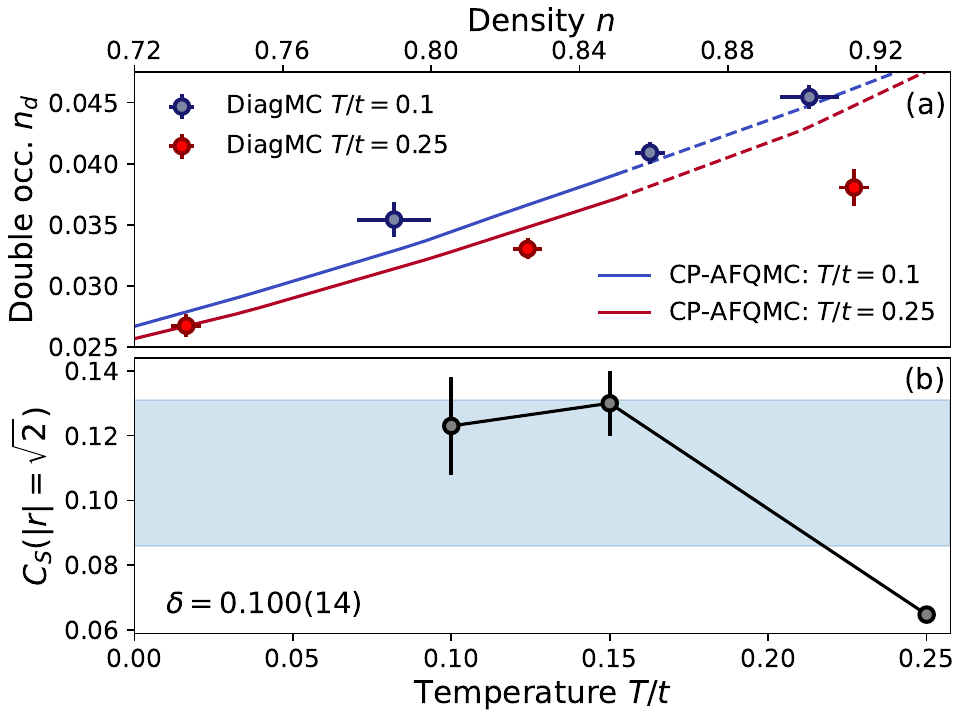}      
\caption{(a) Double occupancy $n_d$ against density $n$ at $U/t=8$ computed by DiagMC and compared with CP-AFQMC data of Ref.\cite{Xu2025}. For densities $n \geq 0.85$, CP-AFQMC could not be directly applied due to poor convergence, and instead a linear interpolation is performed between $n=0.85$ and  half-filling \cite{Xu2025}, shown as the dashed lines. 
(b) 
Temperature dependence of the next-nearest-neighbour spin-spin correlator at doping $\delta = 0.100(14)$ computed by DiagMC and compared with the experimental value at temperature $T/t\lesssim 0.1$ (horizontal band). 
The apparent slight decrease in $C_S(r)$ with cooling at the lowest temperatures is due to a small increase in the doping. However, the doping of all three points remains within the errorbar of the experimental doping $\delta = 0.100(14)$ to which we are comparing.
}    
\label{fig:doublons}
\end{figure}

Given the substantial discrepancy between the CP-AFQMC and experiment/DiagMC data for the spin-spin correlator at distance $d=(1,1)$, it is worthwhile considering possible sources of systematic error.
1) Numerical determination of density: Due to parity-projection during imaging of the atoms, the experiment cannot measure the density directly, and can only measure the singlon density. Reference \cite{Xu2025} therefore used numerical double occupancy data obtained using CP-AFQMC to convert singlon to electron/atomic density using the formula $n = n_s +2n_d$. In Fig.~\ref{fig:doublons}(a), we show a comparison between the density and double occupancy obtained from DiagMC and CP-AFQMC. 
Although a significant discrepancy is observed at $T/t=0.25$, the origin of which requires further investigations, this only leads to a small systematic error in the conversion at this temperature. Meanwhile, excellent agreement is found at the lower temperature $T/t=0.1$. In the supplementary material, we show the singlon density dependence of the spin-spin correlation function at $T/t=0.1$. This suggests that the conversion used in Ref.~\cite{Xu2025} is sufficiently accurate and hence that it cannot account for the discrepancy.
2) Finite size error in CP-AFQMC: The reduced magnitude of the correlations at $T/t=0.1$ could potentially be due to finite size error within the CP-AFQMC calculation. This would explain the discrepancy if CP-AFQMC for a larger system would lead to larger spin correlations, potentially bringing it in-line with the experiment and our TDL DiagMC data. However, at the higher temperature $T/t=0.25$ (see Fig.~\ref{fig:d_11}) the discrepancy between the CP-AFQMC on a $12\times 12$ system and our TDL results is equally sizeable, whereas the DQMC data on an $8 \times 8$ system show good agreement with our results. Moreover, a similar discrepancy was found between DQMC and CP-AFQMC, both performed on a $12 \times 12$ system, at the even higher temperature $T/t=0.33$ \cite{Xu2025}. 
Therefore, while CP-AFQMC is found to be accurate for local and short-range observables, our findings indicate that underestimated systematic error in the CP-AFQMC at distances beyond one lattice spacing is the most likely source of the discrepancy reported in Ref.~\cite{Xu2025}, and shown in Fig.~\ref{fig:d_11}.

Given our agreement with the spin-spin data at $d=(1,1)$ and its substantial increase as the temperature is lowered from $T/t=0.25$ to $T/t=0.1$, a natural question is whether it can be used for a more accurate determination of the experimental temperature. In Fig.~\ref{fig:doublons}(b), we show the temperature dependence of the $d=(1,1)$ correlator at doping $\delta = 0.100(14)$, for which the temperature dependence of spin-spin correlations is strongest, as seen in Fig.~\ref{fig:d_11}. We observe a saturation to the zero temperature regime already at $T/t\approx 0.15$, where it takes a value consistent with the experiment at $T/t\lesssim 0.1$. This means that the spin-spin correlator at this distance does not depend sensitively enough on temperature for it be to used as a thermometer in this regime. 
\begin{figure}[h]
\centering    
\includegraphics[width=0.5\textwidth]{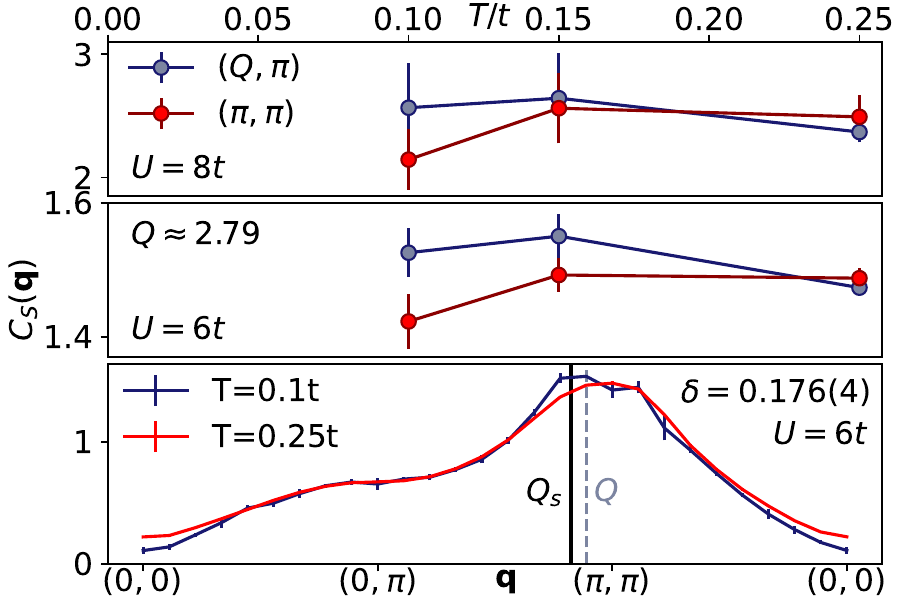}      
\caption{
Temperature dependence of spin correlations $C_S(\mathbf{q})$ in momentum space at the $\mathbf{q}=(\pi,\pi)$ point and $\mathbf{q}=(Q,\pi)$, with $Q\approx 2.79$ nearby the predicted spin-stripe wavevector $Q_S$ (see main text).
}
\label{fig:Sq_vs_T}
\end{figure}
Since long-range correlation functions will be more sensitive to temperature, a better observable for this purpose is likely to be the spin-spin correlator in momentum space $C_S(\mathbf{q})$. 

At moderate to low temperatures, down to $T/t \sim 0.25$, for doping $\delta\simeq0.1$, the momentum-dependence of the spin correlator has a peak at $\mathbf{q}=(\pi,\pi)$ as this region near half-filling is dominated by AFM correlations.
As the system is cooled into the regime of interest, this dominant feature splits along the $k_x$ and $k_y$ directions into incommensurate maxima at $\mathbf{q}=(Q,\pi),(\pi,Q)$ as the predicted spin and charge striped ground state~\cite{Chan2017} is approached. 
The wavelength of these spin-stripes is expected to be related to the doping by $\lambda_s = 2/\delta$, corresponding to $Q_s=(1-\delta)\pi$ \cite{Chan2017, Zhang2022}.
The value of $C_S(\mathbf{q})$ at $\mathbf{q} \sim (Q_s,\pi),(\pi,Q_s)$ ought to be a good choice for a thermometer in this regime since the size of the maxima should depend strongly upon temperature on approach to the stripe order.

Our results for $C_S(\mathbf{q})$ are shown in Fig.\ref{fig:Sq_vs_T}.
At $U/t=6$, the suppression at the $\mathbf{q}=(\pi,\pi)$ point due to the splitting of the AFM peak as the temperature is lowered can be seen clearly, however large statistical error results in too little resolution for the same behaviour to be precisely observed at $U/t=8$.
In addition, the $\mathbf{q}=(Q,\pi)$ peak, whilst distinct, has not yet developed into a prominent feature of $C_S(\mathbf{q})$ at the accessed temperature. 
Upon closer approach towards the stripe order this peak will become more pronounced, potentially making $C_S(\mathbf{q})$ a good choice for thermometry at ultracold temperatures $T/t < 0.1$.

Quantum simulations of fundamental models of condensed matter physics, such as the Hubbard model, are emerging as a crucial tool for discovering exotic quantum states. Recent experimental realizations of the SU$(N)$ Hubbard model~\cite{Taie2022, Pasqualetti2024} have already challenged -- and in some cases surpassed -- state-of-the-art numerical methods, reaching low temperatures with non-trivial magnetic correlations~\cite{Taie2022}. 
For the SU$(2)$ variant, which has a smaller Hilbert space and less severe sign problem in quantum Monte Carlo, classical numerics appeared to hold an advantage — until the work of Xu \textit{et al.}~\cite{Xu2025} achieved a dramatic decrease in temperature.
Our results establish Diagrammatic Monte Carlo as a uniquely powerful tool for benchmarking ultracold-atom quantum simulations of the Hubbard model. The agreement of the controlled-precision calculations with the measurements of Ref.~\cite{Xu2025} validates the simulator’s accuracy in an unprecedented low-temperature regime, opening the door to discoveries of new quantum phases and, potentially, high-temperature superconductivity. At the same time, we identify a critical challenge: conventional thermometry based on the nearest-neighbour \emph{and} next-nearest-neighbour spin–spin correlations \cite{Boll2016} saturates below $T/t \sim 0.15$, limiting its utility. Longer-distance correlations, captured by $C_S(\mathbf{q})$, 
offer a promising path forward for accurate thermometry in correlated regimes near half-filling at these ultralow temperatures.

\begin{acknowledgments}
The authors are grateful to Muqing Xu, Aaron Young, Lev Kendrick, Markus Greiner, and their colleagues at Harvard's Lithium Lab for helpful discussions.
This work was supported by EPSRC through Grant No. EP/X01245X/1. The calculations were performed using King's Computational Research, Engineering and Technology Environment (CREATE).
This work used the ARCHER2 UK National Supercomputing Service (\href{https://www.archer2.ac.uk}{https://www.archer2.ac.uk}) \cite{ARCHER}.

\end{acknowledgments}

\bibliography{biblio}

\end{document}